\documentclass[preprint2]{aastex}

\newcommand{\Msun}{M_{\odot}}
\newcommand{\be}{\begin{equation}}
\newcommand{\ee}{\end{equation}}

{

}

\shorttitle{The density profiles of hot galactic halo gas}
\shortauthors{Hansen \& Sommer-Larsen}
\slugcomment{To be submitted to ApJL}

\begin{document}

\title{The density profiles of hot galactic halo gas}

\author{Steen H. Hansen$^1$ \&
Jesper Sommer-Larsen$^2$}

\affil{$^1$ University of Zurich, Winterthurerstrasse 190,
8057 Zurich, Switzerland}

\affil{$^2$ Dark Cosmology Centre, Niels Bohr Institute, University of Copenhagen, Juliane Maries Vej 30, DK-2100 Copenhagen, Denmark}

%\email{hansen@physik.unizh.ch}



\begin{abstract}
Extended gas haloes around galaxies are a ubiquitous prediction 
of galaxy formation scenarios.
However, the density profiles of this hot halo gas is virtually 
unknown, although various profiles have been suggested
on theoretical grounds.  In order to quantitatively
address the gas profile, we compare galaxies from direct cosmological
simulations with analytical solutions of the underlying gas
equations. We find remarkable agreement between simulations and
theoretical predictions. We present an expression for this gas profile
with a non-trivial dependence on the total mass profile. This expression is
useful when setting up equilibrium galaxy models for numerical experiments.\\
\end{abstract}


%\keywords{}


\section{INTRODUCTION}

Standard galaxy formation scenarios predict that the galaxies have an
extended halo of hot gas.  This hot gas is gravitationally trapped by
the dark matter potential and cools slowly due to thermal
emission~\citep{white91}.

These haloes of hot gas have been observed around massive elliptical
galaxies through their X-ray emission (e.g. Matsushita 2001), and
recently around normal quiescent spiral galaxies~\citep{pedersen2006}.
These observations provide strong support for the standard galaxy
formation scenario.

One missing aspect is the qualitative understanding of the density
profile of these hot haloes. We therefore investigate the density
profiles extracted from virialized galaxies in cosmological
simulations, as well as theoretical predictions based on the
fundamental gas equations. We find remarkable agreement between
simulations and theory. We present an analytical expression for
the gas-profile, which depends non-trivially on the total mass
distribution of both baryons and dark matter.

\section{Theoretical predictions}
Galaxies are dynamically old and should therefore have reached a
quasi-static equilibrium state. In this state there is very little
radial bulk-motion of the hot halo gas, possibly except for the very central
region where cooling is important.

Since the gas has frequent collisions, then we must treat it through
the Navier-Stokes equation (N-S). The N-S equations apply to any
collisional gas or fluid, and are 3 hydrodynamical equations for the
velocity vector.
The galaxies have reached a quasi-static equilibrium state, and then
the N-S  equations simplify considerably since all the
time-dependent terms disappear.

One of the N-S equations is the one related to the radial velocity
component, and if the rotation of the gas is negligible then it
reduces to the normal equation for hydrostatic equilibrium. From this
equation it is straight forward to show that when the mass is
dominated by a spherical distribution of matter, then the spherical
gas density profile is given as a function of the underlying total
matter distribution\footnote{One can show that there are only two
asymptotic solutions to the N-S equations for the gas, one being a
spherical distribution, and one being a disks~\citep{hansenstadel}.}.  One
finds that everywhere throughout the galaxy the
logarithmic derivative of the gas density \be \beta_{gas} \equiv
\frac{\partial {\rm log} \rho_{gas}}{\partial {\rm log} r } \, , \ee
is given as a function of the logarithmic derivative of the total
density, $\beta_{tot}$ and the polytropic index, $\gamma$, which is
related to the derivative of the temperature through \be \gamma = 1 +
\frac{\partial {\rm log} T}{\partial {\rm log} \rho} \, .
\label{eq:gamma}
\ee
One finds the connection
\be
\beta_{gas}  = \kappa \left( \beta_{tot} + 2 \right) \, ,
\label{eq:betagas}
\ee 
where the pre-factor is defined as $\kappa = 1/(\gamma-1)$.  In
general $\gamma$ can be in the range, $1 \le \gamma \le 5/3$, and when
$\gamma=1.5$ we have the connection, $\beta_{gas} = 2 \left(
\beta_{tot} + 2 \right)$.  It is interesting to note that even though,
in general, the solution in the inner region might differ from the
solution in the outer region, as in the hydraulic
jump~\citep{bohr,hansen97}, then it happens that the connection
in eq.~(\ref{eq:betagas}) is the same everywhere~\citep{blois}.

If the mass in a given radial range is dominated by the gas itself,
which may happen in the very inner region, then the solution is
clearly $\beta_{gas} =-2$ ($-4$ or $-6$) if the polytropic index is
$\gamma=1$ ($3/2$ or $5/3$). 

We emphasize that the result in eq.~(\ref{eq:betagas}) is based on the
simplifying assumption that both the gas density profile and the total
mass profiles are power-laws, and therefore eq.~(\ref{eq:betagas})
should only be seen as an approximation.
  
Various other possible behaviours for the gas profile have been
proposed besides the one in eq.~(\ref{eq:betagas}), e.g. that the gas
density profile could follow that of the dark
matter~\citep{santabarbara} \be \beta_{gas} = \beta_{tot} \, .
\label{eq:gastot}
\ee These simulations were considering cluster scales, where cooling is
less relevant (see e.g.~\citep{romeo2006}).  Never the less,
artificially created galaxies used in a range of N-body simulations
(e.g. to investigate aspects of gas-cooling in a controlled 
manner~\citep{kaufmann}),
are routinely constructed under the assumption in
eq.~(\ref{eq:gastot}).  Another suggested behaviour of the gas density
is that is should be virtually independent of the local dark matter
slope, and e.g. follow a $\beta$-profile~\citep{sarazin}, where
$\beta_{gas}$ goes from zero in the center, to $-3\beta$ in the outer
region, which for the normal assumptions imply $\beta_{gas} \approx -2$ in
the outer region.

\section{The code and simulations}
The code used for the simulations is a significantly improved version of
the TreeSPH code, which has been used previously for galaxy formation 
simulations \citep{SLGP}.
The main improvements over the previous version are:
(1) The ``conservative'' entropy 
equation solving scheme suggested by \cite{SH02} has been adopted. 
(2) Non-instantaneous gas recycling and chemical evolution, tracing
10 elements (H, He, C, N, O, Mg, Si, S, Ca and Fe), has been incorporated
in the code following Lia et~al.\ (2002a,b); the algorithm includes 
supernov\ae\ of type II and type Ia, and mass loss from stars of all masses.
(3) Atomic radiative cooling depending both on the metal abundance
of the gas and on the meta--galactic UV field, modeled after \cite{HM96}
is invoked, as well as simplified treatment
of radiative transfer, switching off the UV field where the gas
becomes optically thick to Lyman limit photons on scales of $\sim$ 1~kpc.

The formation and evolution of a total of 15 individual
galaxies, known from previous work to become disk galaxies at $z$=0,
was simulated with the above, significantly improved, TreeSPH code.
At least two different numerical resolutions were used to simulate
each galaxy. Moreover, many of the galaxies were also simulated with different
physical prescriptions for the early ($z\ga4$) star-bursts (and related
SNII driven energy feedback) found previously to be required, in order
to produce realistic disk galaxies. The galaxies were selected to
represent ``field'' galaxies \citep{SLGP}, and span a range of characteristic 
circular speeds of $V_c \sim 100-330$ km/s, and virial masses of 
6x10$^{10}$ to 3x10$^{12} \Msun$.

The galaxies (galaxy DM haloes) were selected from a
cosmological, DM-only simulation of box-length 10 $h^{-1}$Mpc
(comoving), and starting redshift $z_i$=39.  
The adopted cosmology was the flat $\Lambda$CDM model, with
($\Omega_M$, $\Omega_{\Lambda}$)=(0.3,0.7).

Mass and force resolution was increased in Lagrangian regions enclosing the 
galaxies, and in these regions all DM particles were split into a DM particle
and a gas (SPH) particle according to an adopted universal baryon fraction of
$f_b$=0.15, in line with recent estimates. 

In this paper, only results of
two high-resolution simulations of two large disk galaxies will be presented
(see Sommer-Larsen 2006 for details). Each simulation contains about 3x10$^5$ 
SPH+DM particles, and for
the two simulations, $m_{\rm{gas}}$ = $m_*$ =
7.3x10$^5$ and $m_{\rm{DM}}$ = 4.2x10$^6$ $h^{-1}$M$_{\odot}$, and
$\epsilon_{\rm{gas}}$ = $\epsilon_*$ = 380 and 
$\epsilon_{\rm{DM}}$ = 680 $h^{-1}$pc ($h$=0.65). The
gravity softening lengths were fixed in physical coordinates from $z$=6
to $z$=0, and in comoving coordinates at earlier times.
A Kroupa IMF was used in the simulations, and early rapid and 
self-propagating star-formation (sometimes dubbed ``positive feedback'')
was invoked \citep{SLGP}. 

The two galaxies have masses similar to those of
M31 and the Milky Way, with characteristic circular velocities of $V_c =
245$ km/sec and $233$ km/sec respectively. At $z$$\sim$0, each galaxy contains
approximately 10$^5$ dark matter and 10$^5$ gas and star particles
within the virial radius.

At any time there are satellites at various radii in these galaxies,
which produce bumps in the total density profile.  In order to reduce
these bumps, we have co-added 5 frames with 1Gyr spacing,
corresponding to the period z=0.3 to z=0.  No major merging was taking
place during this period.

\begin{figure}[thb]
	\centering
	\includegraphics[angle=0,width=0.49\textwidth]{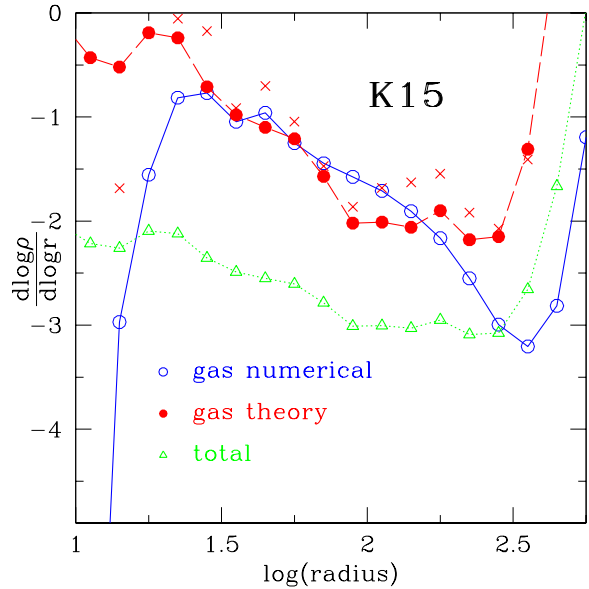}
	\caption{Logarithmic derivative of the 
density profiles of gas and total matter as function
of radius. The solid blue line (open circles) is the simulated gas
density profile (the logarithmic derivative), and the green dotted
line (open triangles) is the simulated total density profile. The red
dashed line (filled circles) is the theoretical prediction from
eq.~(\ref{eq:betagas})
using $\gamma=3/2$, and the red crosses are
using the simulated value of $\gamma (r)$.
This galaxy has $V_c = 245$ km/sec.
Inside log$(r) = 1.4$ the numerical gas density slope is dominated
by the cold high-density gas disk. Outside log$(r)=2.5$ a 
neighbouring galaxy affects the density slope.}
	\label{fig:k15}
\end{figure}

\section{Comparing theory and simulations}

\begin{figure}[bth]
	\centering
	\includegraphics[angle=0,width=0.49\textwidth]{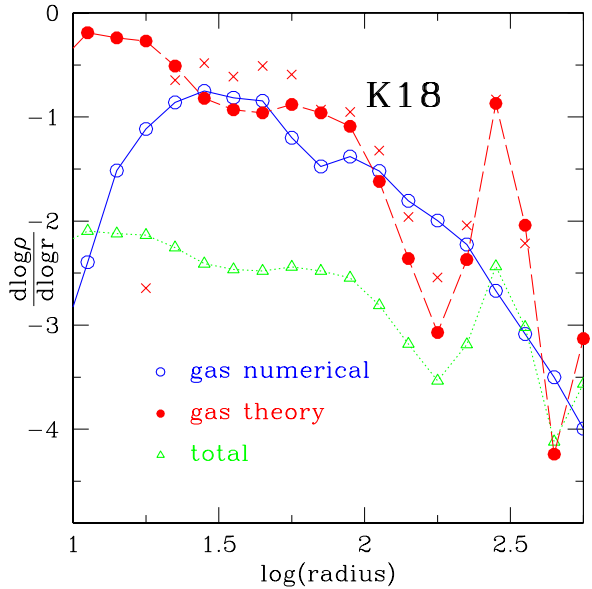}
	\caption{Logarithmic derivative of the 
density profiles of gas and total matter as function
of radius. The solid blue line (open circles) is the simulated gas
density profile (the logarithmic derivative), and the green dotted
line (open triangles) is the simulated total density profile. The red
dashed line (filled circles) is the theoretical prediction from
eq.~(\ref{eq:betagas})  using $\gamma=3/2$, and the red crosses are
using the simulated value of $\gamma (r)$. 
This galaxy has $V_c = 233$ km/sec.
Inside log$(r) = 1.4$ the numerical gas density slope is dominated
by the cold high-density gas disk. Near the virial radius, log$(r) = 2.4$,
the effect of infalling substructure is clearly seen as a bump
in the total density slope.}
	\label{fig:k18}
\end{figure}

In figures~\ref{fig:k15} and~\ref{fig:k18} we plot the logarithmic
derivative of the gas density as a solid, blue line (with open
circles) from the simulation.  We emphasize that many more details are
visible since we consider the derivative of the density profile. In
the very central region the hot gas density profile becomes steep, because the
density is calculated including hot gas particles very near the 
cold, high-density gas disk (i.e., a numerical effect). 
If these near disk hot particles are
excluded, the density slope of the hot gas goes towards zero near the center.

Outside log$(r)$=1.4 the gas slope falls slowly from $-1$ in the
central region, to $-3$ in the outer region near log$(r)$=2.4.  The
virial radius is approximately 250 kpc, corresponding to log$(r)$=2.4.
Further out, for galaxy K15, the presence of a neighbouring galaxy causes
the total and hot gas density profiles to flatten.

Let us now compare this behaviour of the density profile to the
theoretical predictions.  The simulated behaviour of the gas profile
is clearly very different from a beta-profile, which should go from
zero in the central region to roughly $-2$ further out.  It therefore appears
that the hot gas in dynamically old structures, like a galaxy, is not
well described by a $\beta$ profile.

We also plot the total density as a green, dotted line (with open
triangles).  According to the Santa Barbara
comparison~\citep{santabarbara}, the solid, blue line and the green
dotted line could be similar.  It is clear that the behaviour of
derivatives of the total and gas density are very different everywhere
within the virial radius, since the gas does not follow $\beta_{gas} =
\beta_{tot}$, though this is often considered as a realistic initial 
condition for galaxy models.  Our results indicate that such
initial conditions are not in agreement with results for galaxies formed
in $\Lambda CDM$ cosmological simulations.

As red dashed line (with solid circles) we have the prediction from
eq.~(\ref{eq:betagas}). There is impressive agreement for radii
outside the central disk region at log$(r)$=1.4 to the virial
radius. The obvious interpretation is that the hot gas is sufficiently
close to hydrostatic equilibrium.

From a practical point of view, this result gives us a very strong
handle on how to set up initial condition for realistic galaxies for
numerical experiments. One should simply use the total density profile
(typically dominated by the underlying dark matter distribution)
together with eq.~(\ref{eq:betagas}).

For the comparisons above we have used $\gamma=3/2$.  From the
simulations we naturally have $\gamma (r)$ in each radial bin from
eq.~(\ref{eq:gamma}), and we find that $\gamma=3/2$ is a very good
approximation in the entire region considered, namely $1.4 < log(r) <
2.4$. In the figures we also present (as red crosses) the prediction
from eq.~(\ref{eq:betagas}) when using the actual value of $\gamma (r)$,
and we see that there is very little difference from simply using
$\gamma=3/2$

\subsection{Additional aspects}
One seemingly disturbing aspect of eq.~(\ref{eq:betagas}) is that if
the total density slope becomes more shallow than -2, then the gas
density profile gets a positive sign.  We find that it is a remarkable
conspiracy that the total density profile indeed remains as steep as
-2 in the virialized region (see green open triangles in the figures).

In the very central region the gas density is dominated by the cold
gas disk.  The predictions for a disk-profile is different than the
predictions for a spherical distribution in eq.~(\ref{eq:betagas}),
and the disk predictions is approximately $d{\rm log}\rho/d{\rm log}r
\approx -2$, which is in rough agreement with the numerical findings
(see blue, open circles for log$(r) < 1.4$).

It was suggested that the tangential velocity distribution might have
a transition from the inner region to the outer region of the form 
\be
v_{\rm tan} = v_\alpha \, \left( \frac{r}{r_\alpha} \right)^\alpha \,
,
\label{eq:v}
\ee 
where $\alpha$ should go from unity in the inner region, to -2 in
the outer region~\citep{hansenstadel}. We find that the tangential
velocity is well fitted with the shape $v_{\rm tan}= {\rm
exp}(-r^{2.3})$ everywhere in the resolved region. This simply shows,
that the assumption of a power-law for the rotational gas velocity is not good.
Furthermore, the gas is approximately in hydrostatic equilibrium, and
has only a small radial velocity component in the central region where
cooling is most important.  We hope to investigate the details of the
radial velocity and an exponential tangential velocity in the future.

\section{CONCLUSIONS}
Standard galaxy formation scenarios predict that the central cold
gas-disk and stars are surrounded by an extended halo of hot gas.  We
investigate quantitatively the density profile of this gas-halo.  We
find that results of cosmological $\Lambda$CDM N-body/hydrodynamical 
simulations of the formation and evolution of galaxies, and 
analytical solutions to the fundamental gas equations are in
remarkable agreement. We find that the gas density slope (the
logarithmic derivative) is a non-trivial function of the slope of the
total matter, expressed through eq.~(\ref{eq:betagas}).
This equation is useful when constructing realistic galaxies for controlled
numerical experiments.

\acknowledgments 
SHH is supported by the Swiss National Foundation.
The Dark Cosmology Centre is funded by the DNRF.

\end{document}